\documentclass[b5paper,twoside]{jpconf}  
\usepackage{graphicx}

\begin{document}

\title[LAMOST Quasar Survey]{LAMOST Quasar Survey}  

\author[X.-B. Wu]{Xue-Bing Wu\footnote{On behalf of the LAMOST Extragalactic Survey Team}}
\address{Department of Astronomy, Peking University, Beijing 100871, China}

\ead{wuxb@pku.edu.cn}

\begin{abstract}
The main objective of the Chinese LAMOST spectroscopic quasar survey is to discover 0.4 million new quasars from 1 million quasar candidates brighter than the magnitude  limit $i=20.5$ in the next 5 years. This will hopefully provide the largest quasar sample for the further studies of AGN physics and cosmology. The improved quasar selection criteria based on the UKIDSS near-IR and SDSS optical colors are presented, and their advantages in uncovering the missing quasars in the 'quasar redshift desert' are demonstrated. In addition, some recent discoveries of new quasars during the LAMOST commissioning phase are presented.
\end{abstract}

\section{Introduction}
Quasars are interesting objects in the universe since they can be used as 
important tools to probe
the accretion power around supermassive black holes, the intergalactic 
medium, the large scale structure and the cosmic reionization.
The number of quasars has increased
steadily in the past four decades\cite{ric09}. Especially, a large number of quasars have 
been discovered in two recent spectroscopical surveys, namely, the Two-Degree 
Fields (2DF) survey\cite{boy00} and Sloan Digital Sky Survey (SDSS)
\cite{yor00}. 2DF mainly selected lower redshift ($z<2.2$) quasars with UV-excess 
\cite{smi05}, while SDSS adopted
a multi-band optical color selection method for quasars mainly by
excluding the point sources in the stellar locus of the color-color 
diagrams\cite{ric02}. The efficiency
of identifying quasars with redshift between 2.2 and 3 is low in SDSS
\cite{sch07}, because quasars with such redshift usually have 
similar optical 
colors as stars and are thus mostly excluded by the SDSS quasar candidate 
selection algorithm. Therefore, the quasar sample of SDSS is incomplete, and the redshift range
from 2.2 to 3 is also regarded as the 'redshift desert' of quasars because
of the difficulty in identifying quasars within this redshift range.

The Large Sky Area Multi-Object Fibre Spectroscopic Telescope (LAMOST) is
a 4-meter class reflecting Schmidt telescope with 20 square degree field
of view (FOV) and 4000 fibres\cite{su98}.  It is located at the Xinglong Station of
National Astronomical Observatories of Chinese Academy of Sciences. After finishing its main 
construction in 2008, LAMOST has entered the commissioning phase since 2009.
Some test observations have been done in 2009 and 2010. Although
LAMOST has not reach its full ability in the commissioning phase, these
observations already lead to the discovery of some new quasars\cite{huo10}\cite{wu10a}\cite{wu10b}.
The main objectives of the LAMOST quasar survey is to discover 0.4 million new quasars at magnitude limit $i=20.5$ from 1 million quasar candidates selected with the improved selection criteria in the next 5 years. This will hopefully provide the largest quasar sample for the further studies of AGN physics and cosmology in the near future.

\begin{figure}
   \centering
   \includegraphics[width=13cm,height=9.5cm]{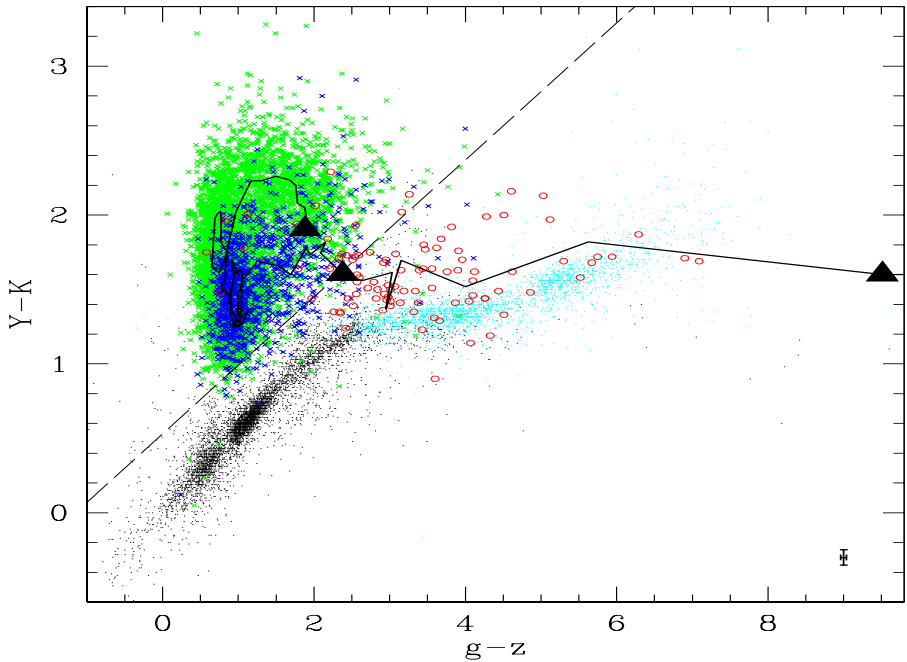}
\caption{The Y-K vs. g-z color-color diagram of the SDSS- UKIDSS quasar and star samples. Black and cyan dots represent stars, while green,blue and red crosses represent quasars with $z<2.2$, $2.2<z<4$ and $z>4$ respectively. Dashed line indicates the quasar selection criterion. Solid curve is derived from the median color-z relation of quasars (see \cite{wuj10}).} 
   \end{figure}

\section{Quasar candidate selection} 
Although quasars in the 'redshift desert' ($2.2<z<3$) have similar optical colors as stars, 
they are usually more luminous than normal stars in the infrared
K-band \cite{war00}. An important way of finding these missing quasars 
has been suggested by using the infrared K-band excess based on the UKIRT
(UK Infrared Telescope) Infrared Deep Sky Survey (UKIDSS) \cite{war00}
\cite{hew06} \cite{mad08}. Recently, based on a 
SDSS-UKIDSS sample of 8498 quasars, Wu \& Jia
proposed to use $Y-K$ vs. $g-z$ diagram to select $z<4$ quasars and use 
$J-K$ vs.$i-Y$ diagram to select $z<5$ quasars\cite{wuj10}. In Fig. 1 we show the  $Y-K$ vs. $g-z$ diagram 
of SDSS-UKIDSS quasar sand stars, as well as the proposed quasar selection criterion\cite{wuj10}, 
$Y-K>0.46(g-z)+0.53$ (when $g$, $z$, $Y$, $K$ are all in Vega
magnitude) or $Y-K>0.46(g-z)+0.82$ (when $g$, $z$ are in AB magnitude and $Y$, $K$ are in Vega
magnitude).
Recent spectroscopic observations made
by us \cite{wu10a}\cite{wu10b}\cite{wu11} have demonstrated the high success rate in identifying $2.2<z<3$ quasars from 
the quasar candidates selected by these new quasar selection criteria, especially when the variability information is combined.
Beside using the SDSS-UKIDSS selection method, we also
included some quasar candidates from other catalogs\cite{ric09}\cite{bov11}, as well as
the additional candidates selected from the SVM method by the NAOC team \cite{gao08}
for preparing the targets of the LAMOST quasar survey.

 \begin{figure}
   \centering
   \includegraphics[width=13cm,height=12cm]{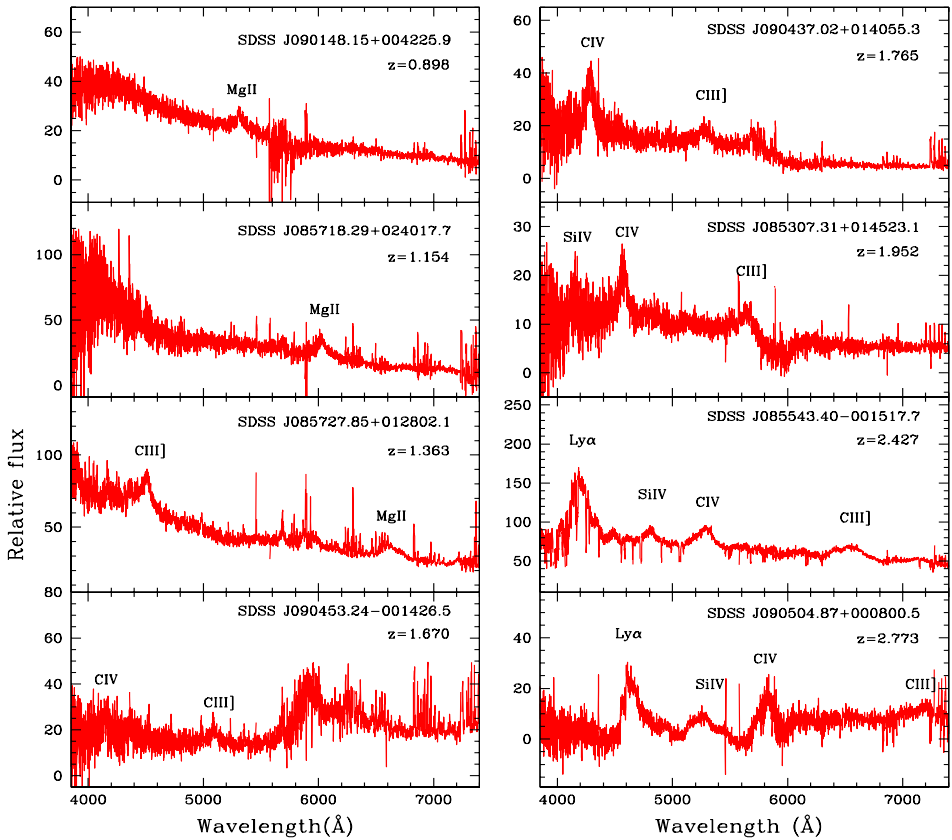}
\caption{Spectra of eight new quasars discovered by LAMOST in one shot (see \cite{wu10b}).} 
   \end{figure}

\section {New quasars discovered by LAMOST}
LAMOST entered its commissioning phase in 2009.
In the winter of 2009, we have selected several extragalactic fields for the 
LAMOST commissioning observations.On December 18, LAMOST made the spectroscopic observations on one field centered at
RA=$08^h58^m08.2^s$, Dec=$01^o32'29.7''$   with 
the exposure time of 30 minutes and the spectral resolution of $R\sim1000$. Eight new quasars are discovered (\cite{wu10a}\cite{wu10b}; See Fig. 2 for their LAMOST spectra). 
We noticed that two of eight new quasars have redshifts larger than 2.2.  
These quasars in the 'redshift desert' are very difficult
to be identified because of their similar optical colors as stars.
However, they can be recovered by using the combination of optical and near-IR colors. 
Although LAMOST met some problems during the commissioning 
observations, we were still able to identify many other known SDSS 
quasars in this field, with $i$ magnitudes from 16.24 to 19.10 and redshifts 
from 0.297 to 4.512. The discovery of new quasars 
supports the idea that by combining the UKIDSS near-IR colors with the
SDSS optical colors we are able to efficiently recover the missing quasars in
the SDSS spectroscopic survey even at the magnitude limit $i<19.1$.

\section{Discussion}
A complete quasar sample is very important to the construction of the quasar 
luminosity function and study the cosmological evolution of quasars.
However, because $2.2<z<3$ quasars have similar 
optical colors as normal stars, it is very difficult for find them in the optical 
quasar surveys. The low efficiency of finding quasars in the redshift desert 
($z$ from 2.2 to 3) has led to obvious incompleteness of SDSS quasar sample 
in this redshift range and serious problems in 
constructing the luminosity function for quasars around the redshift peak 
(between 2 and 3) of quasar activity \cite{ric06}\cite{jia06}. 
Therefore, recovering
these missing quasars with the improved quasar selection criteria will become an important task in the future quasar surveys.
We hope that great
progress will be made in improving the capability of LAMOST spectroscopy before its normal sky survey in 2012. 
As long as LAMOST 
can reach its designed capability after the commissioning phase, we expect to obtain 0.4million quasars
from 1million candidates in the LAMOST quasar survey in the next 5 years. This will form the largest quasar sample and will
undoubtedly play a leading role in the future quasar study. 

\section*{Acknowledgment}  The work is supported by an NSFC grant (No. 11033001) and the 973 program (No. 2007CB815405). LAMOST is a National Major Scientific Project built by the Chinese Academy of Sciences, and is operated and managed by the NAOC.

\end{document}